\documentclass[sigconf]{acmart}

\usepackage{booktabs} 

\usepackage{amsthm}

\usepackage{latexsym}
\usepackage{amsmath}
\usepackage{amssymb}
\usepackage{graphicx}
\usepackage{subcaption}
\usepackage{amsmath}
\usepackage{graphicx}
\usepackage{caption}
\usepackage{subcaption}
\usepackage{float}
\usepackage[vlined,algoruled,titlenumbered,noend]{algorithm2e}
\usepackage{hyperref}
\usepackage{ctable}
\hypersetup{
	colorlinks   = true,
       linkcolor = red,
            urlcolor  = blue,
            citecolor = blue
}
\usepackage{url}

\copyrightyear{2017} 
\acmYear{2017} 
\setcopyright{acmcopyright}
\acmConference{ICTIR '17}{October 1-4, 2017}{Amsterdam, Netherlands}

\begin{document}
\title{Characterizing and Predicting Supply-side Engagement on Crowd-contributed Video Sharing Platforms}

\renewcommand{\shorttitle}{Supply-side Engagement on Crowd-contributed Video Sharing Platforms}

\author{Rishabh Mehrotra}
\affiliation{
  \streetaddress{University College London}
  \city{London} 
  \state{United Kingdom} 
}
\email{r.mehrotra@cs.ucl.ac.uk}

\author{Prasanta Bhattacharya}
\affiliation{
  \streetaddress{National University of Singapore}
  \city{Singapore}
}
\email{prasanta@comp.nus.edu.sg}

\settopmatter{printacmref=false, printfolios=false}

\begin{abstract}
Video sharing and entertainment websites have rapidly grown in popularity and now constitute some of the most visited websites on the Internet. Despite the active user engagement on these online video-sharing platforms, most of recent research on online media platforms have restricted themselves to networking based social media sites, like Facebook or Twitter. We depart from previous studies in the online media space that have focused exclusively on demand-side user engagement, by modeling the supply-side of the crowd-contributed videos on this platform. The current study is among the first to perform a large-scale empirical study using longitudinal video upload data from a large online video platform. The modeling and subsequent prediction of video uploads is made complicated by the heterogeneity of video types (e.g. popular vs. niche video genres), and the inherent time trend effects associated with media uploads. We identify distinct genre-clusters from our dataset and employ a self-exciting Hawkes point-process model on each of these clusters to fully specify and estimate the video upload process. Additionally, we go beyond prediction to disentangle potential factors that govern user engagement and determine the video upload rates, which improves our analysis with additional explanatory power. Our findings show that using a relatively parsimonious point-process model, we are able to achieve higher model fit, and predict video uploads to the platform with a higher accuracy than competing models. The findings from this study can benefit platform owners in better understanding how their supply-side users engage with their site over time. We also offer a robust method for performing media upload prediction that is likely to be generalizable across media platforms which demonstrate similar temporal and genre-level heterogeneity.
\end{abstract}


\maketitle

\section{Introduction}
Video delivery and sharing platforms have witnessed exponential growth in recent times, with websites such as Youtube, Netflix and XVideos.com featuring among most frequently visited websites worldwide\footnote{http://www.alexa.com/topsites/countries;1/US}. However, what sets these websites apart from other social media sites such as Facebook, Twitter and  Reddit is the sheer amount of data traffic that passes through their servers, as almost all content on these sites is available as streaming videos\footnote{http://www.cnet.com/news/netflix-youtube-gobble-up-half-of-internet-traffic/}. Further, the average amount of time spent by any user on these platforms is reported to be much higher than that spent on any other leading social media sites\footnote{http://www.comscore.com/Insights/Press-Releases/2014/2/comScore-Releases-January-2014-US-Online-Video-Rankings}.

Despite the obvious importance of such video distribution and sharing platforms, academic research on such websites have been relatively lagging, with most choosing to focus on network-oriented sites instead \cite{leskovec2008planetary} \cite{leskovec2008microscopic}. The extant research in the area of user-generated-content (UGC) too has been primarily focused on textual and pictorial content \cite{leskovec2009meme} \cite{kumar2010structure} \cite{cha2007tube}, and on issues related to UGC based predictions in the real world \cite{asur2010predicting} \cite{o2010tweets} \cite{luo2013social}. Most importantly, the few studies that have explicitly looked at video-based content, investigate demand-side research questions about video downloads, and consumption patterns \cite{figueiredo2013prediction, cha2007tube}. However, what lags in these previous investigations on the topic is an understanding of how the supply-side engagement develops on these platforms.

User engagement has been coined as the "emotional, cognitive and behavioral connection that exists between a user and a resource" \cite{attfield2011towards}. While most users engage with these platforms as content-consumers, their engagement is sustained by a steady upload of videos from a relatively smaller sample of video uploaders who engage actively with the platform as content producers. As a result, it is imperative for such platforms to predict and understand user engagement so as to constantly attract content contributors by keeping them users engaged (for the purpose of this work, by "user" we imply users who contribute content, i.e., content producers). 

In this current study, we depart from existing work by explicitly modeling the supply side engagement pattern of these video uploaders across different video genres. Specifically, we introduce a point process model to specify the content generation process and provide a mechanism for predicting future upload volumes and disentangling the contributing factors. For our empirical analysis, we use data from a large-scale adult entertainment website which is ranked among the top two most frequently visited adult entertainment sites on the Internet. Our dataset comprises information about the uploaded video (e.g. title, descriptions etc.) as well as user-generated tags associated with each upload. In order to uncover genre-level engagement patterns from the upload data, we first perform a clustering analysis, forming association clusters based on the co-occurrence of video tags which are selected by the users while uploading the video to the platform. Thus, videos which have a common set of associated tags are inducted into the same video cluster, implying that each video cluster qualitatively represents a specific genre or taste category.

Video uploads demonstrate a significant rate-heterogeneity depending on the specific video genre as is evident from the distinct upload patterns in each of the 4 illustrative clusters shown in Figure \ref{fig-uploadpatterns}. Drawing on this insight, we perform predictive modeling on each of the clusters individually to generate insights about the upload process for the specific video genre. For our model estimations, we employ a parametric self-exciting process model, also termed as a Hawkes process model in literature \cite{hawkes1971spectra}. Such models provide an elegant and parsimonious extension to the popular Poisson model, by incorporating the history of events into consideration. Self-exciting models are ideal candidates for fitting multi-spell events with bursty traffic where there are infrequent spikes in frequency followed by periods of mean-reveal when the frequency retreats to its mean value. In our current study, we apply the Hawkes model to each of the identified genres, also termed as clusters, and obtain parameter estimates that we later use to make predictions. We show that our model fits the data better than comparable variants of the Poisson model that have been used in recent research on Hawkes models. Moreover, our model provides lowest average prediction error spanning different splits of the training and test data, as compared to other baseline models.

In addition to predictive accuracy, we attempt to disentangle the
effect of various factors contributing to the video upload process,
and investigate which among them were the driving contributors to
video uploads, spanning different clusters. Identifying the contributing factors helps in better understanding the psyche of users engaging with the platform as content contributors. We assert that the engagement driving factor derives from three major sources viz. self-reinforcement, popularity of other videos in the genre, and other exogenous events. We perform an explanatory analysis to estimate what fraction of the clustering in each genre can be attributed to each of these three sources. Thus, drawing on our self-excitation model, we are not only able to make accurate predictions of video uploads, but are also able to explain the source for these upload intensities across different video categories.

We contend that our study is among the first to go beyond consumer side view to analyze the supply side of videos generated in a real-world setting. While our empirical analysis leverages data from a large adult entertainment platform, the models and methods we use can easily be adapted to other video streaming platforms without any loss of generality. Specifically, we offer the following three contributions in the current paper: First, while several studies on UGC in general, and videos in particular have focused on modeling demand-side user engagement patterns (i.e. content consumption), this is among the first studies to analyze the user-generated supply side nature of these video distribution platforms. 
Second, we leverage a Hawkes point-process model to provide a robust predictive model which outperforms other comparable baselines that do not take into account the self-exciting nature of video uploads. Third, we go beyond prediction, to uncover potential factors that determine the video upload rate. This improves the analysis with additional explanatory power. We contend that these findings will increase our understanding of video-based UGC production on online entertainment platforms, and will aid platform owners in better understanding how their content producers engage with with the platform by producing content that has high genre-level and temporal heterogeneity.

\vspace{-4mm}
\section{Related Work}
We review some past work looking at user engagement on online platforms. Since, in the current study, we focus our attention at specifically studying videos uploads, we also provide a brief review of studies that look at video distributions.\\

\noindent\textbf{User Engagement in Online Platforms:}\\
In the online industry, web analytics is used to understand how users engage with a site and includes metrics such as click-through rate, time spent on a site (dwell time), page views, return rates and number of
users. These metrics, referred to as engagement metrics, assess users' depth of engagement with a site. Although they cannot explain why users engage with a site, they have been used as proxy for online user engagement. Existing studies of user engagement with a web service can broadly be classified into three groups. First, several studies focused
on analysis of user behavior. Some of them discovered the
relationship between goal success and system reuse \cite{white2010modeling}. Some others concerned behavioral patterns of users (e.g. models of web sites with respect to user behavior (e.g.,
w.r.t. multitasking user behavior \cite{lehmann2013online} or w.r.t. popularity, activity, and loyalty among users \cite{lehmann2012models}).  Second, some studies focused on the prediction of future changes in user engagement. Prediction of how a user switches (no switch, persistent switch, or oscillating behavior) between different online systems during 26 weeks was studied in \cite{white2010modeling}. Third, there are papers devoted to user engagement as an evaluation metric in online controlled experiments \cite{drutsa2015practical,kim2015ir}.

While most existing work has focused on analyzing user behavior and engagement, in this work we go a step beyond: in addition to making predictions, we aim at understanding user engagement with a crowd-contributed video upload website by dis-entangling the different factors associated with user's psyche while contributing content.

\noindent\textbf{Video-based User Generated Content:}\\
The emergence of online communication and, in particular, on social media has dramatically increased online engagement and word-of-mouth (WOM), or user-generated content (UGC), on such platforms. These WOM interactions, mostly textual, have been used to predict movie revenues,and television success \cite{rui2011designing,chintagunta2010effects,asur2010predicting}, election outcomes \cite{metaxas2012social}, product sales \cite{goh2013social} and even firm equity values \cite{luo2013social}. There have been related research on modeling the emergence and growth of such text-based user generated content as well \cite{xu2012modeling,toubia2013intrinsic}.\\


\noindent\textbf{Self-exciting Point Processes:}\\
Our approach in this work is based on the Hawkes Process which is a type of self-exciting point-process model. Such point-process models have been popularly used in recent studies to model natural phenomena like wildfire assessments \cite{peng2003applications},  spiking in brain-waves \cite{reynaud2013spike}, financial settings\cite{filimonov2013apparent} and even online check-ins on a social media site\cite{cho2014and}.

\begin{table}[t!]
\centering
\resizebox{0.5\textwidth}{!} 
{
	\begin{tabular}{|c|c|c|}
	\specialrule{.1em}{.05em}{.05em}
	\textbf{Metadata} & \textbf{Description} & \textbf{Example}\\
	\specialrule{.1em}{.05em}{.05em}
	upload\_date & 	Day when the video was uploaded	 & 4//30//2011\\
	\hline
	title	 & Title of the video & 	"Tea party at Dick's house"\\
	\hline
	channels & 	List of the video's tags & 	['Tea', 'Spoon', 'Sugar']\\
	\hline
	description	 & Description of the video & 	"What a spoon !"\\
	\hline
	nb\_views	 & Number of times the video has been displayed & 	69\\
	\hline
	nb\_votes & 	Number of users who voted for or against this video & 	42\\
	\hline
	nb\_comments	 & Number of comments posted on this video & 	666\\
	\hline
	runtime & 	Length of the video in seconds & 	4815\\
	\hline
	uploader	 & Anonymized identifier of the uploader's username	 & 6f60cbef5b891f80\\
	\specialrule{.1em}{.05em}{.05em}
	\end{tabular}
}
\caption{The metadata associated with each uploaded video on the website \cite{mazieres2014deep}.\label{metadata}}
\vspace{-8mm}
\end{table}

\section{Dataset Description}
For our empirical analysis, we use data from a large-scale adult video sharing site\footnote{http://xhamster.com/} \cite{mazieres2014deep}. Such genre-based video sharing platforms like 9Gag.tv, XVideo.com etc. are unique in that they provide a fertile venue to study both genre-level heterogeneity (i.e. heterogeneity in uploader's video tastes), as well as temporal perturbations (i.e. periods of temporal clustering). Our dataset comprises an exhaustive collection of metadata from all videos published on a large scale adult video platform, since its creation in  April 2007, up until February 2013, totaling over 800,000 videos from over 85000 uploaders. Table \ref{metadata} describes the associated metadata which consists of an anonymized uploader identifier, video upload date and time, list of uploader contributed video-tags for the uploaded videos, and video popularity cues (e.g. number of views, comments etc).

The content on such adult sites is often arranged in the form of diverse categories or tags for ease of access and use. While uploading any new video, the uploader is given the option of selecting tags from a set of existing tags. Often a video is tagged with multiple related tags. It is to be noted that the tags appear to form some clusters based on tag associativity, an insight we leverage later for better predictions.

\begin{figure*}[!]
\centering
\includegraphics[width=1.0\textwidth, height = 8cm]{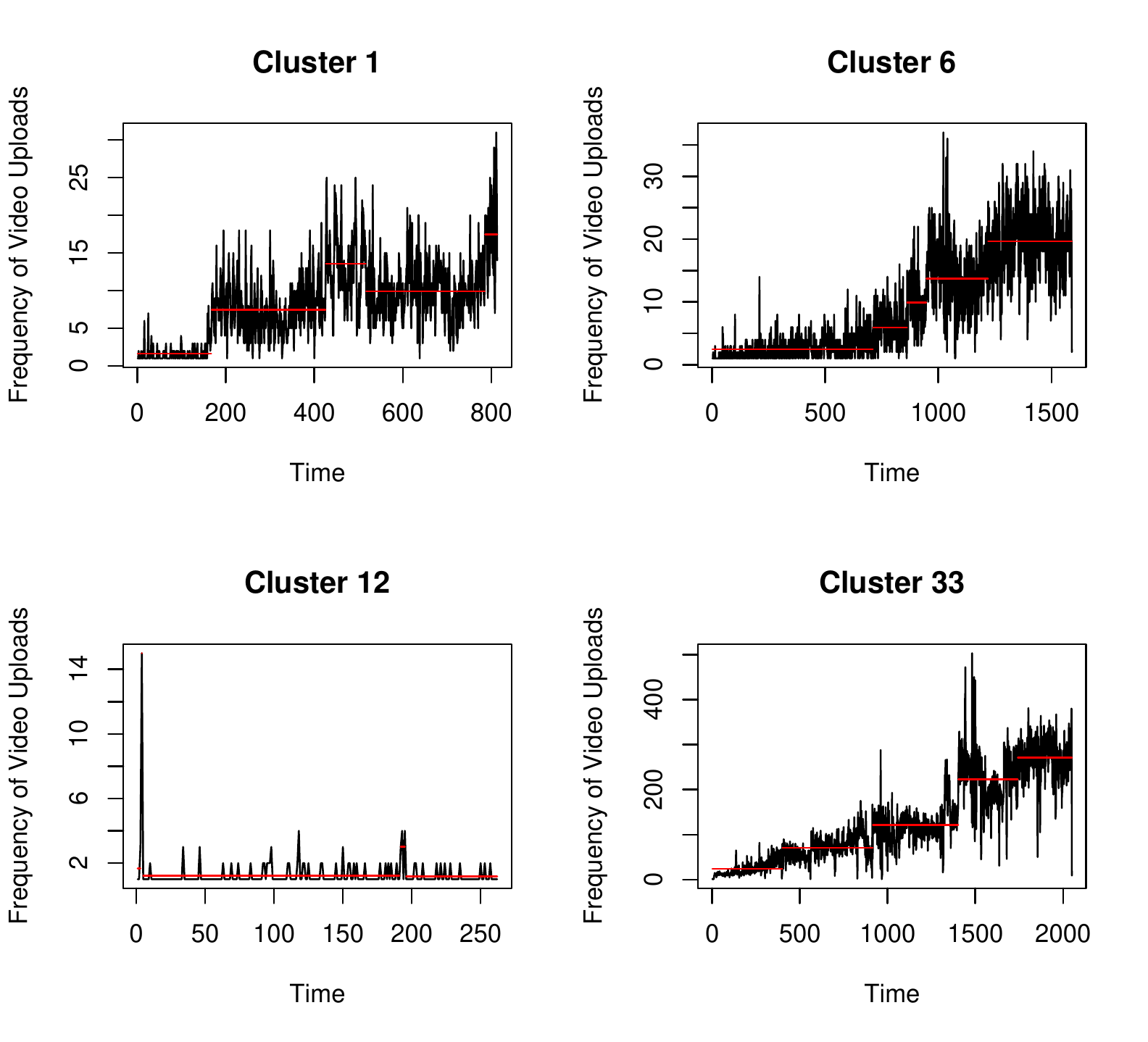}
\vspace{-3mm}
\caption{Variation in upload patterns across different clusters. \label{fig-uploadpatterns}}
\vspace{-4mm}
\end{figure*}

\section{Modeling Video Uploads}

Our aim in this research is to characterize the video upload process in such crowd-contributed adult video ecosystems so as to be able to uncover emerging genres in video uploads and predict future upload volumes for each of these themes, while at the same time explaining what factors might be generating the observed volumes. A multitude of factors influence the supply-side process of user contributions on such platforms. The inherent heterogeneity of content as reflected by the abundance of user-selected tags and categories poses interesting challenges in analyzing the content upload process. Additionally, the temporal variability introduced by seasonal trends and popularities form yet another aspect of the upload process.

We first explain the process of extracting different clusters of videos based on their tag associativity, following which, we present a Hawkes Process to model the upload process and predict future uploads.


\subsection{Graph-based Tag Cluster Formation}
Almost any web site that provides means for sharing user-generated multimedia content has tagging functionalities to let users annotate the material that they want to share. The tags are then used to retrieve the uploaded content, and to ease browsing and exploration of these collections, e.g. using tag clouds. These tags also provide additional contextual and semantic information by which users can organize and access shared media content. While uploading new media content, users typically associate their media with those tags which can potentially explain the content better to prospective viewers. Each media can thus be associated with different tags thereby forming a complex network of inter-tag relationships. When analyzing tag associations, it is often observed that some sets of tags co-occur together in a large proportions, suggesting that, together, they can be viewed as a high-level genre or category. While some of these genres or categories (e.g.funny cat videos) are widely popular, others are very niche (e.g. black hole videos) with often a particular set of users responsible for most of the content generation and consumption in such categories. Often the content generation rate in niche videos are triggered by some external events (e.g. a recent movie related to Astronomy triggers an increased interest in black hole videos). This necessitates the need for segregating such categories of videos because the way a popular category gets flooded with user generated content varies drastically from the content generation process in such niche categories. 

While videos do not have explicit category/cluster labels, we propose to make use of the associated tag information to uncover the various clusters underlying our adult entertainment videos. To this end, we formulate the tag associations in videos in a graph based setting. Given a set of videos along with their tag associations, we build a complete graph $G_{V} = (T, E, w)$, whose nodes $T$ are the set of all tags associated with the set of videos $V$, and whose $E$ edges are weighted by the tag-tag affinities. The weighting function $w$ is a tag affinity function $w : E \rightarrow \mathbb{I}$ where $\mathbb{I}$  is the set of integers. For each pair of tags, the edge weight is defined as
\begin{equation}
w(t_1,t_2) = |\{v\}| \  s.t. \ t_1\in tags(v) \ \& \ t_2\in tags(v)
\end{equation}
i.e.,  the total number of videos in which these two tags co-occur. Overall, the graph $G_{V}$ describes the tag-affinity network for the set of videos $V$.

We define the video clusters as the set of vertex-partitions induced by the connected components of the graph $G_V$. The rationale is to drop weak edges, i.e., low tag-affinity and to build clusters on the basis of the strong edges, i.e. with high tag affinity, which identify the related tag pairs. The algorithm performs two steps:
\begin{itemize}
\item [1] \textbf{Graph Pruning:} given the graph $G_V$ all the edges $e \in E$ whose weight is smaller than a given threshold, that is $w(t_1,t_2) < \eta$, are removed, thus obtaining a pruned graph $G'_V$. 
\item [2] \textbf{Connected Components:} in the second step, the connected components of the pruned graph $G'_V$ are extracted.
\end{itemize}
Such connected components identify the genre-clusters of related tags which are returned by the algorithm.

\subsection{Genre-Cluster Analysis}
After constructing the tag-tag affinity graph for the contributed video tags and applying a vertex partitioning algorithm, we were able to uncover a total of 37 genre or category-based clusters from the entire dataset, such that videos within each cluster had similar tags. 
The average number of videos within each cluster was 19460, with the minimum and maximum number of videos in any cluster being 337 and 264509 respectively, for cluster numbers 12 and 33.

\subsection{Modeling Video Uploads as a Hawkes\\Process}
\label{hawkes}
The \emph{weighted connected components} of the tag affinity graph, as defined above, serve as the genre-clusters which represent the different types of video categories one usually observes on online video sharing sites. As is evident from the distinct upload patterns in each of the 4 illustrative clusters Fig. \ref{fig-uploadpatterns}, video uploads demonstrate significant rate-heterogeneity depending on the specific video genre. Such cluster specific heterogeneity warrants the need to perform predictive modeling on each of the clusters individually to generate cluster-specific insights. We treat each such genre-clusters as a separate process and we adopt the parametric Hawkes process for each genre-cluster to model the cluster specific upload process.\\

\subsubsection{Hawkes Process}
A point process $N$ is a random measure on a completely separable metric space $S$ that takes values
on $N \cup \lbrace\infty\rbrace$. In our case, a convenient way to view a realization of $N$ is that of a list of times $t_1, t_2, ...., t_n$ at which events $1, 2, ...n$ occur. A point process is typically characterized by prescribing its conditional intensity $\lambda(t)$, which represents the infinitesimal rate at which events are expected to occur around a particular time t, given the history of the process up to t, $H_t = {t_i : t_i < t}$ \cite{ogata1988statistical} Thus, in a point process, $N(t)$ counts the number of points (i.e., occurrences of events) in $(-\infty, t]$, and the conditional intensity function $\lambda(t|H_t)$ denotes the expected instantaneous rate of future events at timestamp $t$ depending on $H_t$, the history of events preceding $t$. An important example of a point process is the Poisson process, which always has a deterministic conditional intensity $\lambda(t)$. We say that a point process N is self-exciting if
\begin{equation}
Cov[N(t_1,t_2), N(t_2,t_3)] > 0
\end{equation}
for any $t_1<t_2<t_3$ . This means that if an event occurs, a successive event becomes more likely to occur locally in time and space. This is, however, not true for a Poisson process which has independent increments, hence $Cov[N(t_1,t_2), N(t_2,t_3)]  = 0$.

The Hawkes process is a specific class of self- or mutually-exciting point process models \cite{hawkes1971spectra}. A univariate Hawkes process $\lbrace N(t)\rbrace$ is defined by its intensity function
\begin{equation}
\lambda(t) = \mu(t) + \int_{-\infty}^t \kappa(t-s) dN(s)
\end{equation}
where $\mu : \Re \rightarrow \Re^+$ is a deterministic base intensity, $\kappa : \Re^+ \rightarrow R^+$ is a kernel function expressing the positive influence of past events on the current value of the intensity process. The process is well known for its self-exciting property, which refers to the phenomenon that the occurrence of one event in the past increases the probability of events happening in the future. Such a self-exciting property can either exist between every pair of events, as assumed in a normal univariate Hawkes process, or only exist between limited pairs of events.\\

\subsubsection{Modeling Video Uploads}
Each genre-cluster obtained via the connected components of the tag affinity graph is treated as a separate Hawkes Process. Each video upload in the genre-cluster is treated as an event in the given cluster specific point process. We model the intensity of video upload events involving a cluster $c$ at time $t$ as follows:
\begin{equation}
\lambda_{c}(t) = \mu_c + \sum_{p:t_p<t} g_c(t-t_p)
\end{equation}
This intensity function can be interpreted as a rate at which video-uploads in a cluster occur. The summation in the second term is over all the events (i.e. uploads) that have happened up to time $t$. $\mu_c$ describes the background rate of event occurrence that is time-independent, whereas the second term describes the self-excitation part, so that a video upload in the past increases the probability of observing another upload in the (near) future. We will use a two-parameter family for the self-excitation term:
\begin{equation}
g_c(t-t_p) = \beta_c exp(-w_c(t-t_p))
\end{equation}
where $\beta_c$ describes the weight of the self-excitation term (compared to the background rate), while $w_c$ describes the decay rate of the excitation. Intuitively, the decay term captures the notion that more recent upload events are more important.

Overall, each genre-cluster is defined by three sets of parameters of the Hawkes Process: $<\mu,\beta,w>$ representing the upload process characterized by the particular cluster. The estimates of these parameters were obtained by minimizing the negative of the log likelihood function \cite{peng2002multi}.

\section{Disentangling the Contributing Factors}
As can be seen in Fig \ref{fig-uploadpatterns}, the average upload patterns across the different clusters are not static, but rather vary significantly with time and across clusters. Such a dynamic pattern can be attributed to the different factors which trigger users to contribute new media. In this section, we describe three primary mechanisms to describe the video upload process: (i) self-reinforcing behavior of users, (ii) trend-burst influence, also called here as the "popularity effect," and (iii) other exogenous factors. We are especially interested in disentangling the individual contributions of these three effects from the overall cumulative effect. The rich information in the data including the user information and the content popularity information allows us to construct a fine-grained model of the strength of the effect of one event on the other. We next describe the process of finding such relations, a technique which we use to uncover the relative contributions of these three different effects.\\

\subsection{Inferring Correlations between Events}
\label{corr}
A Hawkes process model provides us with the flexibility to characterize the relationships between two events (e.g. between successive uploads, as in this study). We can infer the strength of the ties between two events by examining the intensity function for a given event which further allows us to infer the likelihood that the event was triggered by a specific historical event. A current event can potentially be triggered by any of the historical events. We use a probabilistic measure (introduced in \cite{cho2014and}) described below, to model the strength of tie between i and j. For the given process (representing a specific genre-cluster c), the probability that the $j^{th}$ event is triggered by the $i-th$ event can be expressed as below:
\begin{equation}
p^c_{i\rightarrow j} = \frac{g_c(t_j-t_i)}{\mu_c + \sum_{p:t_p<t_j} g_c(t_j-t_p)}
\end{equation}
From the parameter values learnt in section \ref{hawkes} above, we can calculate the above probability based on the cluster specific $<\mu_c,\beta_c,w_c>$ values. Since we are interested in correlation of points for a given process (cluster) and not the correlation across different processes (clusters), we assume each process (cluster) has its own parameters which we estimate in isolation from each other.

\subsection{Contributing Factors}
\label{factors}
Our goal here is disentangling and analyzing the different factors along with their individual contributions towards explaining the content creation volume in any particular genre-cluster. We consider three major factors that govern the content generation behavior observed in crowd-contributed websites. We describe each of the three in detail below. The user-views and comment information associated with each video serves as a proxy to estimate the relative popularity of the different videos, and along with the uploader's information, it provides us with the necessary equipment to tackle the disentanglement objective. 
Individual contributions from each of these factors allow us to construct a fine-grained model of the strength of the effect of one on another. While the ground truth cause for each video upload is unknown, we offer various ways to quantitatively test the validity of our factors.\\

\begin{figure*}[t]
\centering
\includegraphics[width=\textwidth, height = 5cm]{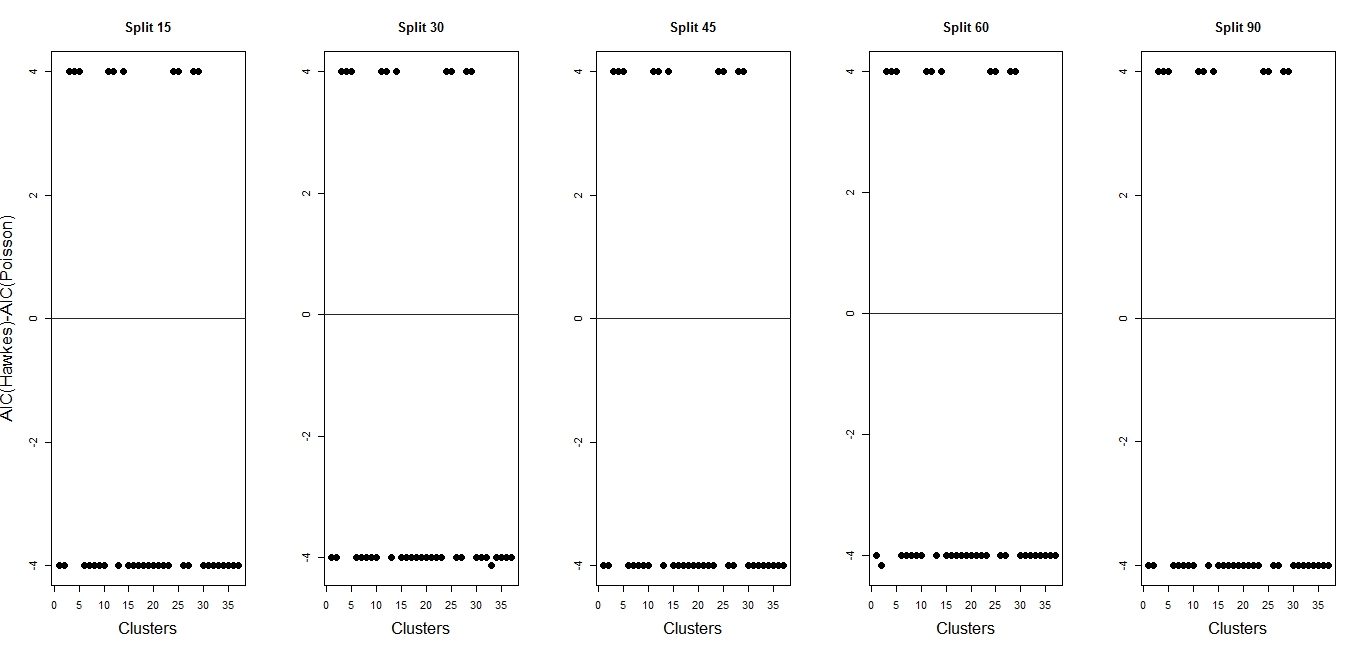}
\vspace{-8mm}
\caption{Comparison of model fit statistics: the difference in AIC scores among the proposed Hawkes process model and the Poisson model is plotted for each of the genre-clusters.\label{fig-modelFit}}
\vspace{-4mm}
\end{figure*}

\subsubsection{Self-reinforcing Behavior}
Often, users exhibit strong predictable behaviors in terms of their affinity towards a particular genre-cluster. Quite frequently, a particular user consistently uploads videos belonging to the same genre. Such a repetitive and self-reinforcing behavior observed on the platform is indeed a major factor governing the proportion of content in any particular cluster. A user who has already uploaded in a particular cluster is more likely to display behavioral consistency and upload again soon in this cluster and, conversely, a paucity of uploads strongly predicts fewer uploads in the future. This self-reinforcing tendency is measured using the event correlation equation described in subsection \ref{corr} by summing over upload events i and j that were initiated by a particular user. More specifically, we define the self-reinforcing score for a particular genre-cluster $c$ as follows:
\begin{equation}
S_{self} = \frac{\sum_{t_i < t_j}p^c_{i\rightarrow j}\mathbb{I}_{u_i=u_j}}{\sum_{t_i < t_j}p^c_{i\rightarrow j}}
\end{equation}
where $\mathbb{I}$ is the indicator function which equals one if the uploader $u_i$ is same as uploader $u_j$ and $i$ \& $j$ correspond to the upload event taking place in genre-cluster $c$.\\

\subsubsection{Popularity Effect}
The perceived \emph{popularity} of an already uploaded content often lends a sense of validation to a prospective uploader who might use this information to decide whether or not to upload content of that genre \cite{goes2014popularity}. Based on this intuition, we postulate that users are more likely to upload content if they perceive that the content they are uploading will be well-received by the content consumers. To incorporate this effect, we make use of a user specific \emph{popularity effect} which is our second major factor governing the proportion of content in a particular cluster.

The average popularity of uploaded content would differ for different users - some users would have a relatively high popularity average while some would have a relatively low popularity average based on their past uploads. Indeed, different users have different notions of baseline popularity thresholds which are often impacted by how popular user's past uploads were. We quantify the popularity of any uploaded content in terms of the number of views and the number of comments it has received (popularity score for a video upload event i is notated as $\psi_{i}$). We model user specific popularity threshold by averaging over the user' past popularity score and postulate that past uploads by other users in a particular cluster which are, on average, more popular than the focal user's average popularity score, the \textit{popularity effect}, will positively impact the uploader's decision to upload content to that cluster. We define the Popularity Effect score for any particular genre-cluster $c$ as follows:
\begin{equation}
S_{pop} = \frac{\sum_{t_i < t_j}p^c_{i\rightarrow j}\mathbb{I}_{\psi_{i} > \psi_{avg_{u_j}}}}{\sum_{t_i < t_j}p^c_{i\rightarrow j}}
\end{equation}
where $\psi_{i}$ denotes the popularity of video event $i$, $\psi_{avg_{u_j}}$ denotes the average popularity score for user $u_{j}$ and $i$, $j$ correspond to the upload events taking place in the genre-cluster $c$. Thus, the term $\mathbb{I}_{\psi_{i} > \psi_{avg_{u_j}}}$ activates all past video upload events wherein the video $i$ was more popular than user $u_j$'s average popularity score. This effect models the increased likelihood of a upload event taking place in light of past popularity of content from the same cluster.\\

\noindent\textbf{Exogenous Effect}\\
If a upload event is not explained by either of the effects above, we consider it to be caused by some external (exogenous) factors. Indeed, other factors like website streaming quality, consumer demand, ease of content creation process, etc. might contribute towards explaining the observation that some genre-cluster have much more upload events than others. While modeling these effects individually provides stronger cues and insights about the overall upload process, for the current study, we accumulate them together as \textit{Exogenous Effects}. The score for the exogenous effect is calculated based on the scores of the two main factors described earlier.
\begin{equation}
S_{exo} = 1 - S_{self} - S_{pop}
\end{equation}

We make use of the above mentioned scores to evaluate the impact of these different factors towards guiding genre-cluster level upload behavior and present detailed results in Section \ref{results2}.

\begin{table*}[t!]
\centering
{
	\begin{tabular}{|c|c|c|c|c|c|c|}
	\specialrule{.1em}{.05em}{.05em}
	\textbf{Splits} & \textbf{Hawkes} & \textbf{PC-NHPP} & \textbf{NHPP-D}  & \textbf{No Cluster}  & \textbf{Baseline 4}\\
	\specialrule{.1em}{.05em}{.05em}
	15 & \textbf{31.99} & 32.00 & 33.50 & 49.23  & 44.26\\
	\hline
	30 &  \textbf{27.76} & 31.39 & 30.96  & 50.12 & 43.06\\
	\hline
	45 &  \textbf{35.25} & 40.60 & 38.16 & 49.89 & 43.78\\
	\hline
	60 &  \textbf{37.43} & 37.58 & 38.63256  & 51.22 & 44.02 \\
	\hline
	90 &  \textbf{39.30} & 41.90 & 42.89 & 51.45 & 44.79\\
	\specialrule{.1em}{.05em}{.05em}
	\end{tabular}
\caption{Predictive Analysis: the error in predicting total number of video uploads within a future window of 2 weeks. The splits highlight the number of training days used to construct the model.\label{tab:tab2}}
}
\vspace{-4mm}
\end{table*}

\begin{figure*}[t]
\centering
\includegraphics[width=\textwidth, height=5cm]{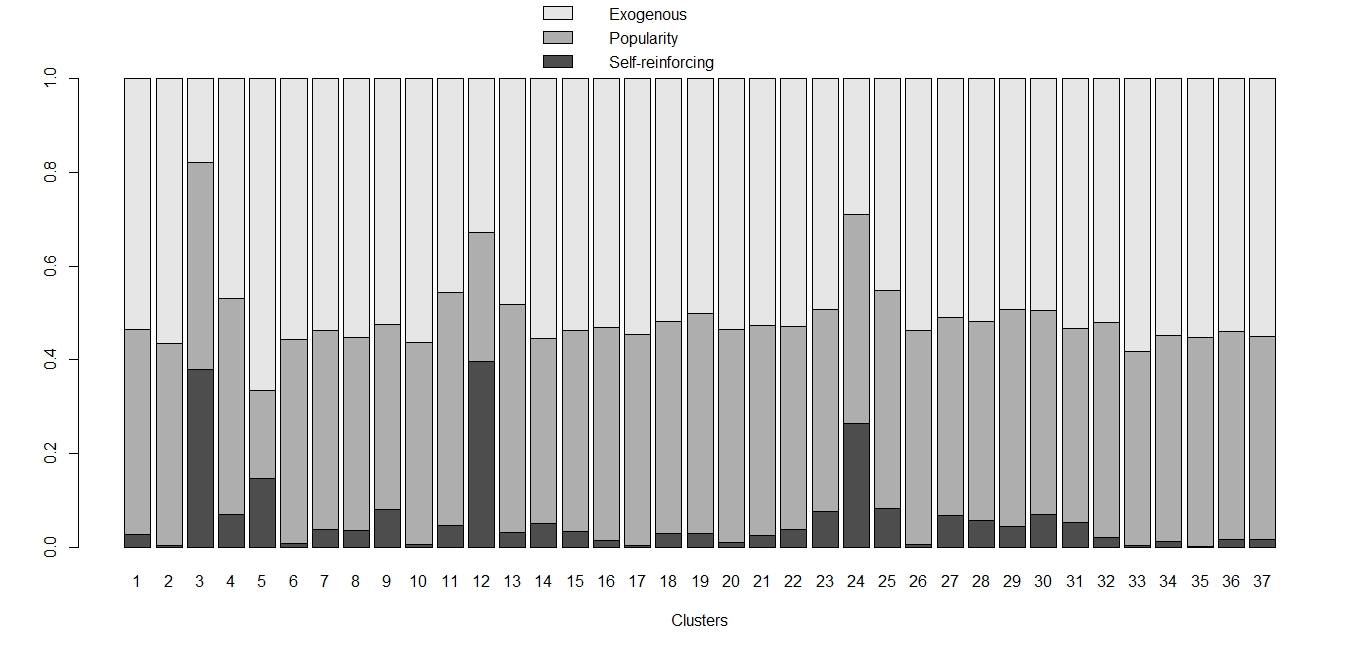}
\vspace{-9mm}
\caption{Relative impact of the three contributing factors across the different genre-clusters. \label{fig-factors}}
\vspace{-4mm}
\end{figure*}

\section{Experimental Evaluation}
We next evaluate our performance in modeling and predicting video uploads via comparing goodness of fit and prediction errors of the proposed Hawkes process model with a number of baselines as described below.

\subsection{Baselines}
\label{baselines}
We compare the Hawkes process model to several other baselines based on nonhomogeneous Poisson processes(NHPP) as described below.\\

\noindent\textbf{Baseline 1:  piecewise-constant NHPP (PC-NHPP)}\\
Content uploads rates for each cluster follow a cluster specific background rate: some popular genre-clusters generally notice much frequent uploads than other niche genre-clusters. We fit a piecewise-constant nonhomogeneous Poisson processes(NHPP) for each cluster and use the results as our first baseline.\\

\noindent\textbf{Baseline 2: NHPP with drifting (NHPP-D)}\\
We define the rate function as a linear function of time. On many genre-clusters, the video uploads become more frequent as time elapsed from the first upload. A sudden video upload suddenly sparks interest among consumers which excites uploaders to upload more content to this particular genre-cluster. To model this, we use a rate function defined as $\lambda(t) = \mu t + b$, where $\mu$ is the cluster specific the base rate.\\

\noindent\textbf{Baseline 3: Hawkes process with no clusters}\\
In this baseline, we consider all historic events ignoring the cluster assignments and model a single Hawkes process model on the entire data.\\

\noindent\textbf{Baseline 4: ARIMA modeling}\\
As a final baseline, we also estimate a time series model based on an autoregressive integrated moving average (ARIMA) specification. We implement a series of model specifications for each of the clusters by varying the autoregressive order, the moving average order and the degree of differencing. We finally select the model with the lowest AIC criterion and use it to make forecasts. The results are shown in Table \ref{tab:tab2}.

\subsection{Model Selection}
For every genre-cluster, we fit the data to a Hawkes process using MLE and evaluate the goodness of fit compared against other baseline approaches (see Section \ref{baselines} for the list of baselines). For evaluation we use the AIC score \cite{burnham2002model}, which has been widely used for model selection. In addition to maximizing likelihood, AIC also penalizes models with large number of parameters to discourage overfitting. The model with the smallest score is chosen from the candidates.

Figure \ref{fig-modelFit} in the previous page shows a comparison of the model fit between our proposed Hawkes model and a homogeneous Poisson model, that is popularly employed to model count- or rate-related data. The difference in Akaike Information Criterion (AIC) between the Hawkes and Poisson model forms our key criteria for comparison. It is clear that across different sample sizes, denoted by the split values, our model shows superior model fit as compared to its Poisson counterpart. This is evidenced by a lower AIC value across most clusters for a given data size.

\vspace{-2mm}
\subsection{Predicting Video Uploads}
To additionally evaluate the performance of the proposed Hawkes Process based model for modeling the video upload process, we evaluate our model on the task of video upload prediction. For all genre-clusters, we segregate the data into two components, training set and testing set and fit a separate Hawkes Process on the training data and perform MLE to obtain estimates of the model parameters. Using these parameters, we intend to predict the number of videos that would be uploaded in a given future time frame. With the estimated parameters, the rate function at time $t$ can be computed based on the history up to time $t$ and the parameters estimated from the training set. The number of events (video uploads) between time interval $t$ and $t + \delta t$ can be computed using the counting process as below $(\delta t > 0)$:
\begin{equation}
N(t + \delta t) - N(t) = \int_t^{t+\delta t} \lambda (\tau) d\tau
\end{equation}

In our experiments, we focus on predicting the number of videos uploaded in the time frame of two weeks. For this, we make use of a number of training-testing splits. The training splits consists of data from the past 15 days, 30 days, 45 days, 60 days and 90 days. It is to be noted that the upload process at the web scale is highly susceptible to seasonal trends such that a holiday season experiences a surge in the number of videos uploaded as compared to a more monotone season. Hence, modeling dependencies in the upload process based on a longer historical past would add noise to the training phase. 

We report results from our predictive analyses where the performance of the Hawkes model in predicting total number of video uploads to the site within a future window of 2 weeks is analyzed. As described in the previous section, we also run two comparable baseline models viz. a piecewise constant NHPP , and NHPP with drifting\cite{DBLP:conf/aaai/ChoSG14}. Our results as illustrated in Table \ref{tab:tab2} show that the Hawkes model outperforms both of the Poisson based baseline models across all sample sizes. The lower the prediction error, the better the model at predicting the upload volume. We find that our model is able to predict video uploads to the site with prediction error rates lowest among comparable models used in recent studies.\\

\noindent\textbf{Impact of considering clusters:}\\
As can be seen from the predictive results, cluster specific Hawkes process model performs better than the No Cluster Hawkes process model. Indeed this highlightsthe importance of modeling each cluster separately via a cluster specific model parameters. \\

\noindent\textbf{Benefits over Time Series Model:}\\
The performance of the ARIMA baseline is worse than that of the proposed model which highlights the fact that simple time series based models aren't generic enough to incorporate variations while Hawkes process is able to better model the temporal variations in the upload process.\\

\vspace{-4mm}
\section{Impact of Contributing Factors}
\label{results2}
One of the major goals of this study was to also disentangle the effect of various factors contributing to the video upload process, and investigate which among them were the driving contributors to video uploads, spanning different clusters. As mentioned in Section \ref{factors}, we hypothesize that there are three major factors that contribute to temporal clustering viz. self-reinforcing behavior, popularity effect, and unobserved exogenous factors.

Figure \ref{fig-factors} demonstrates the distribution of the self-reinforcing score, popularity effect score and exogenous effect score for each of our 37 clusters.  We find evidence for both self-reinforcing behavior and popularity effect within each cluster. However, quite interestingly, we do find variation in the relative proportion of these two scores across the clusters. For instance, while most clusters report a higher popularity effect as compared to the self-reinforcing scores,  clusters 3,12 and 24 report a higher than average self-reinforcing effect. Further, clusters 3 and 12 report even higher self-reinforcing scores than popularity effect. Taken together, these results hint of a strong genre-level dependency that exists on these video platforms, and while popularity of other videos in a genre is a leading driver for most genre of videos, this is not necessarily true for all genres. Our findings uncover this interesting interplay between video genres and the factors contributing to increased uploads. 

\vspace{-4mm}
\section{Conclusion}
The current study is among the first to fully characterize, explain and predict supply side engagement patterns on a large-scale online video sharing platform. We uncover significant user- and genre- level heterogeneities in online video uploads, and propose a parametric self-exciting point process model for modeling the same, after controlling for genre-level heterogeneity and temporal perturbations. We demonstrate a higher model fit as compared to a homogeneous-rate Poisson model, and also make more accurate predictions than comparable baseline models. Additionally, beyond predicting uploads, we also discuss possible reasons for the high-clustering behavior observed in our dataset. We posit that such clustering effects in uploader engagement could be the result of self-reinforcing behavior of the uploads, or due to popularity influence of other videos of the same genre, or even due to other exogenous factors that are not captured in our dataset. Based on our conditional intensity modeling, we are able to successfully disentangle the above three causes for the observed video clustering, thereby providing evidence that both self-reinforcement as well as the popularity effect have a strong role to play in producing the upload pattern displayed by our data. By providing a parsimonious model that combines the benefit of predictive modeling with strong explanatory power, and a unique dataset that allows us to investigate both genre-level and temporal heterogeneities, we offer researchers, policy makers, economists as well as platform owners with useful insights about the supply-side behavior of video-based entertainment sharing platforms. With the growing importance and accessibility of such platforms, we believe that such computational approaches would be increasingly beneficial in managing and securing such ecosystems.


\bibliographystyle{ACM-Reference-Format}
\bibliography{sigproc-less}

\end{document}